\documentclass{aastex}
\usepackage{emulateapj5}
\usepackage{apjfonts}

\newcommand{\kms}{${\rm km \ s^{-1}}$}

\newcommand{\fluxconv}{${\rm erg \ cm^{-2} \ s^{-1} \ (DN \ s^{-1})^{-1}}$}
\newcommand{\lam}{$\lambda$}
\newcommand{\lamlam}{$\lambda \lambda$}
\newcommand{\lya}{Ly$\alpha$}
\newcommand{\ha}{H$\alpha$}
\newcommand{\hb}{H$\beta$}
\newcommand{\civ}{\ion{C}{4}}
\newcommand{\oii}{[\ion{O}{2}]}
\newcommand{\oiii}{[\ion{O}{3}]}
\newcommand{\nii}{[\ion{N}{2}]}

\lefthead{TELFER, KRISS, \& TSVETANOV}
\righthead{SEARCH FOR EXTENDED LINE EMISSION FROM BALQSOS}

\begin{document}
\submitted{Accepted for publication in the November issue of the Astronomical Journal}
\title{A Search for Extended Line Emission from BALQSOs$^1$}

\author{Randal C. Telfer\altaffilmark{2}, 
Gerard A. Kriss\altaffilmark{2,3}, and Zlatan Tsvetanov\altaffilmark{2}}

\altaffiltext{1}{Based on observations with the NASA/ESA 
{\it Hubble Space Telescope}, obtained at the Space Telescope Science 
Institute, which is operated by the Association of Universities for Research 
in Astronomy, Inc., under NASA contract NAS5-26555.
These observations are associated with proposal ID 7892.}
\altaffiltext{2}{Center for Astrophysical Sciences, Johns Hopkins University, 
Baltimore, MD, 21218-2686}
\altaffiltext{3}{Space Telescope Science Institute, 3700 San Martin Drive, 
Baltimore, MD, 21218}

\begin{abstract}
We have obtained {\it HST} NICMOS snapshot images of ten BALQSOs and four 
radio-quiet non-BALQSOs
to search for extended narrow-line regions.  For each object we obtained an image
in the F160W filter and a narrow-band image in either \ha, \hb, or \oii.
We find no detections of extended narrow-line emission in the sample.  We generate
simulated images of QSOs with extended narrow-line regions to place limits on the
amount of flux that may be present, and we conclude that BALQSOs cannot be more
than several times brighter in extended emission than typical type 2 AGN.
Spatially unresolved \ha\ emission is detected in all nine objects observed in this
line, which we attribute
to broad \ha\ emission.  The five detected BALQSOs have derived \ha\
equivalent widths that are larger than those of the four non-BALQSOs by a factor
of $\sim$2 on average.  This may point to an intrinsic difference between
the rest-frame optical spectral properties of BALQSOs and non-BALQSOs.

\end{abstract}


\section{INTRODUCTION}
Broad absorption line QSOs (BALQSOs) exhibit broad absorption 
troughs blue-shifted from the peaks of the corresponding broad emission 
lines, typically including \lya\ \lam 1216,
\ion{C}{4} \lam 1549, \ion{Si}{4} \lam 1397, \ion{N}{5} \lam 1240, and 
\ion{O}{6} \lam 1034.
They represent a small fraction of optically-selected QSOs, about 9\%,
though the inferred fraction increases to about 12\% 
when the luminosities are corrected for flux lost in the absorption 
troughs \citep{foea90}.  Additional effects such as scattering attenuation of
the continuum \citep{good97} or a tendency for BAL lines of sight to lie
at large inclination angles with respect to the accretion disk \citep{krvo98} would
make BALQSOs even more underrepresented in flux-limited surveys, and thus the 
true population fraction of BALQSOs could be significantly greater than 12\%,
perhaps as large as 50\%.
One of the most basic and important problems concerning BALQSOs is the
population fraction and how it relates to the location and geometry of 
the absorbing material.  At one extreme, the BALQSOs could be a separate 
class of QSOs with a large, nearly unity, covering fraction of absorbing 
clouds.  Alternatively, in a unified view, the population fraction of BALQSOs 
is related to the typical covering fraction of the absorbing gas.
Through a detailed study of emission and absorption line
profiles of BALQSOs, \citet{hkm93}
conclude that the covering fraction of absorbing material is generally
$\lesssim 0.2$.  However, the corrections to the observed population fraction
suggested by \citet{good97} and \citet{krvo98} would imply larger covering fractions.
If the typical covering fraction is indeed much less than unity, 
then it seems plausible that BALQSOs and radio-quiet non-BALQSOs belong to the same
population and differ only in their orientations 
relative to the observer, analogous to unified models of Seyfert 1 and 2 galaxies.
Thus understanding the location and 
geometry of the absorbing gas in a BALQSO could be an important step in 
developing a full picture of the structure of all QSOs.

\citet{weea91} favor a geometry in which clouds 
ablated from a torus
surrounding the central source are accelerated into an outflowing wind.
Lines of sight lying near the plane of the torus for a given
object would result in its classification as a BALQSO.  One could also 
envision a model in which the high-velocity outflow was directed
along the axis of the torus so that BALQSOs are viewed instead pole-on.
Spectropolarimetric observations with the Keck 10-m telescope
\citep{gomi95,coea95}, however,
favor a geometry in which the absorbers are in a flattened
configuration around the nuclear region associated with, or just outside, the
broad emission-line region (BELR), consistent with the \citet{weea91} model.  
The suggested geometry for BALQSOs strongly resembles the
obscuration, reflection, orientation model that provides a unified view of 
Seyfert 1 and 2 galaxies \citep{anto93}, although recent evidence suggests that
some Seyfert 2 galaxies are intrinsically distinct from Seyfert 1s \citep{mgt98}.
In the unified model, Seyfert 1s are observed pole-on with a direct view of the 
central regions;
the line of sight to Seyfert 2s intercepts the obscuring torus, but 
electron scattering provides a view of the nuclear regions.
A corollary of this geometrical model is that shadowing by the torus can
also collimate the ionizing radiation into cones that illuminate gas in the 
surrounding galaxy.  In Seyfert 2s the illuminated gas shows extended, 
bipolar structures projected onto the sky, while in Seyfert 1s the structures 
are expected to be more symmetric and compact \citep{mwt96,scki96}.
If this geometrical picture of BALQSOs is correct, then 
BALQSOs, similar to Seyfert 2s,
are preferentially viewed near the plane of the obscuring torus.  If
QSOs possess extended narrow-line regions (NLRs) collimated by the torus as seen 
in Seyfert 2s, these extended regions will preferentially have larger projections
on the plane of the sky than non-BALQSOs.

A significant difference between BALQSOs and Seyfert 2s is that 
in BALQSOs our line of sight to the nucleus and BELR is relatively clear 
\citep{weea91}.
A completely opaque torus along the line of sight, however, is not necessary
to collimate the radiation.  All that is required is material that
is opaque to {\it ionizing} radiation.
Some Seyfert 1s, notably NGC~4151 and NGC~3516, show strong, broad UV
absorption lines \citep{peea81,ulbo83},
opaque Lyman limits \citep{krea92,krea96}, and extended, bipolar NLRs 
\citep{evea93,mwp92}.
\citet{evea93} and \citet{krea95,krea96} suggest
that the optically thick neutral hydrogen in these galaxies collimates the
ionizing radiation.  The same principal can work in BALQSOs.

Extended narrow-line emission from QSOs has been readily observed and studied
at lower redshift ($z < 1$), but has been mostly confined to radio-loud QSOs.
The presence
of extended \oiii\ emission has been found to be preferentially associated with 
steep-spectrum radio sources \citep{book84, bpo85, stma87}.  \citet{duea94} used
integral field spectroscopy to study the extended emission in greater detail 
and found that the morphology and velocity structure of the ionized gas was quite
irregular and complex.  \citet{hbf96} detected extended \oii\ \lam 3727 emission 
around several radio-loud QSOs.

Following the analogy to Seyfert 2 galaxies, we undertook a snapshot
program with {\it HST} NICMOS to search for extended line emission
that might be associated with an extended NLR in BALQSOs.  No extended
emission was detected in our program.  In \S\ref{sec:obs} and 
\S\ref{sec:analysis} we discuss our observations and analysis, and in
\S\ref{sec:discuss} we discuss the implications of the lack of 
extended emission in our sample.

\section{OBSERVATIONS AND DATA REDUCTION\label{sec:obs}}
We obtained images of 14 QSOs from 1998 March to November as an
{\it HST} NICMOS snapshot program.  Our full sample of 28 QSOs
was chosen from \citet{weea91} and
includes 21 BALQSOs and 7 radio-quiet non-BALQSOs.  
Objects were selected by redshift
such that the narrow emission line of \oiii\ \lam 5007 or \lam 4959 (none observed), 
\oii\ \lamlam 3726, 3729 (4 objects observed), \ha\ \lam 6563 (9 objects observed), 
or \hb\ \lam 4861 (1 object observed) was shifted
to a wavelength within the bandpass of one of the narrow-band filter / camera
combinations available on NICMOS.  All data were acquired in MULTIACCUM
mode.  Each object was imaged in the narrow band for 703.94~s using the
Step64 exposure time sequence and in the F160W filter with the same camera
for 71.93~s using the Step8 exposure time sequence.  The observation
date and filters used for each object are listed in
Table~\ref{ta:observations}.  Unfortunately, the original redshift assigned to the 
BALQSO Q~0021-0213,
$z = 2.293$, which would have placed the \oiii\ emission line in the narrow
band pass, was incorrect.  The revised redshift, $z = 2.348$ \citep{hfc95}, 
places the line outside the band pass
of F164N, and we do not consider this object as part of our observed sample.

All of the images were processed with IRAF using the STSDAS calibration 
pipeline task {\it calnica} along with the {\it biaseq} and {\it pedsky} 
tasks to remove the 
non-linear readout-to-readout bias levels and background pedestal,
respectively.  Some residual shading was evident in the NIC 2 images, which
was removed with the IRAF task {\it background}, fitting the background 
perpendicular to the fast read direction.

We identified a potential problem with the CRIDCALC step in the {\it calnica} 
pipeline that results in underestimates of the errors for MULTIACCUM data.  
To produce the final science and error images, {\it calnica} fits a line to the 
counts vs.\ time data provided by the many MULTIACCUM images and derives the 
final count rate and error per pixel based on this fit.
For data that are read-noise dominated, this is a valid approximation.  
However, the algorithm does not take into account
the fact that the Poissonian errors from read to read in the signal from
any sources are correlated, and
thus underestimates the errors in regions of high signal.  We
generated our own error images using the following formula:
\begin{equation}
{\it error} = \frac{\sqrt{\frac{|{\it image}| * {\rm EXPTIME} * 
{\rm ADCGAIN}}{\rm FLATFILE} + {\rm NOISFILE}^{2}}}
{{\rm EXPTIME} * {\rm ADCGAIN}},
\end{equation}
where {\it image} is the final science image produced by {\it calnica} in
units of ${\rm DN \ sec^{-1}}$, FLATFILE and NOISFILE are the 
particular flatfield image and RMS read noise image in electrons used by
the pipeline, EXPTIME is the exposure time in seconds, and ADCGAIN is
the analog-to-digital conversion gain in $e^{-} \ {\rm DN}^{-1}$.  This 
method overestimates the read noise contribution, since this will in general be 
smaller than the noise per read represented by the NOISFILE due to the multiple
readouts used in MULTIACCUM mode.
We do not include the error from the dark current contribution.  
In order to check that ignoring the dark current in this calculation is justifiable, we
estimate the typical dark current contribution to our images using typical values for
the dark current contribution as described in \citet{skbe97}.
The dark current 
consists of three basic components:  the true linear dark current, amplifier glow from 
infrared radiation emitted by the readout amplifiers, and shading due to time dependence
of the bias level during readout.  The shading component is noiseless.
The linear dark current is only $\sim 0.05 \ e^{-} \ {\rm s^{-1} \ pixel^{-1}}$, 
so even with a 700~s exposure, the error contribution from the linear dark current is 
only $\sim 6 \ e^{-} \ {\rm pixel^{-1}}$ and is thus negligible compared to
the read noise, which is typically $\sim 30-40 \ e^{-} \ {\rm pixel^{-1}}$.  
The largest error contribution
from the dark current is from amplifier glow, which can contribute 
2--3 ${\rm DN \ pixel^{-1}}$ per read
to the signal.  The gain for our data is 5.0 $e^{-} \ {\rm DN}^{-1}$ for cameras 1 and
2, 6.5 $e^{-} \ {\rm DN}^{-1}$ for camera 3.  Given that there are 20 readouts in
the Step64 exposure time sequence, and assuming ${\rm 3 \ DN \ pixel^{-1}}$ 
per read from amplifier glow and a gain of 5.0, the amplifier glow would contribute 
$20 \times 3 \times 5 = 300 \ e^{-} \ {\rm pixel^{-1}}$
to the signal in the final readout, for an error contribution of
$\sim 17 \ e^{-} \ {\rm pixel^{-1}}$.
The contribution is thus not insignificant but still small compared to the read noise
that we include,
and we ignore it for simplicity.  Overall we have found
that the error images produced by this method yield a good approximation of the error
in regions of low signal and, by design, a much better representation of the 
Poissonian errors.  We thus consider it to be a more reliable indicator of the true 
uncertainty for our particular data sets and use it in our analysis.

\section{ANALYSIS\label{sec:analysis}}

We use two different methods to determine the presence and magnitude
of any emission-line flux.  In the first method, we use the broad-band 
image to determine 
the continuum contribution to the count rate in the narrow-band image in order to
calculate the net emission-line flux.  This method ignores the spatial 
distribution of the line emission.  In the second method, we measure the 
residual count rate in PSF-subtracted images.  To quantify limits on the 
amount of extended line emission from these residuals, we create models of an 
extended NLR combined with a point source as well as PSF-subtracted models to 
compare with the real data.

In both of these methods we utilize model PSFs created with 
the {\it HST} PSF generation program Tiny Tim \citep{krst93}
version 5.0.  All the PSFs are calculated for the
observed position on the camera array and for the appropriate observation date.
The PSFs are subsampled by a factor of 10 for better accuracy
in sub-pixel shifting, used for the PSF subtraction.
For simplicity, we use monochromatic PSFs, using for the representative wavelength 
the effective wavelength of the bandpass for the narrow-band PSFs and 1600~nm
for the broad-band PSFs.  Although in general one might not expect a 
monochromatic PSF to be a very good match for F160W data, we use the
broad-band PSFs only to obtain aperture correction factors, and our
tests show that for our apertures these factors vary by $\lesssim 1$\%
throughout the bandpass ($\sim {\rm 1400-1800 \ nm}$), so they are suitable
for this purpose.

Using aperture photometry, we measure the brightness of each QSO in the 
broad-band and narrow-band images.  We choose aperture radii of an integral 
number of pixels such that the aperture encloses at least 90\%
of the light based on model PSFs.  The resulting apertures enclose the
Airy disc and the first bright Airy ring and have radii between $0\farcs 5$
and $0\farcs 8$ in angular extent,
depending on the camera and filter being used.  This size aperture is
suitable since it is large enough to be
insensitive to minor variations in the PSF and to enclose a significant
amount of the extended emission, if it is present, but not so large as
to significantly increase the uncertainty of the measurement due to
additional background and read noise.  These count rates are then
multiplied by an appropriate aperture correction factor as calculated from
the model PSFs.  Table~\ref{ta:fluxes} lists the
raw and corrected count rates with errors as obtained by this method.  
The errors attributed to the measurements are simply the quadrature sums of 
the errors of each pixel included in the aperture, calculated as described 
in \S\ref{sec:obs}.  We also list the F160W magnitudes of each object in AB 
mags.

To derive estimates of the emission-line flux from these data, 
we use the STSDAS package {\it synphot} to calculate the expected ratio of broad-band
to narrow-band continuum count rates for each object assuming
a power-law continuum $F_{\nu} \propto \nu ^{-0.3 \pm 0.6}$.
The power-law index and RMS deviation were estimated by inspection from
the statistical data of \citet{frea91}, 
which covers the continuum wavelength range of interest.  This allows us
to scale the broad-band fluxes to remove the continuum contribution in the
narrow band.  We use {\it synphot} to shift a model line profile to the proper 
wavelength and calculate conversion factors from net emission-line count 
rates to flux for each object.  For \oii\ and \oiii\ we model the line as 
a Gaussian with a FWHM of 400 \kms.  \ha\ and \hb\ are both modeled by the 
broad \hb\ profile extracted from the \citet{frea91} composite.
(\ha\ is outside the wavelength coverage of the composite.)  As applied to \ha, this
model profile does not include any contribution from the narrow lines 
\nii\ \lamlam 6548, 6583, but in broad-line objects these lines make a very small 
contribution to the emission-line flux compared to broad \ha, as can be seen for
example in \citet{jabr91}.  The resulting conversion factors are shown in column~5
of Table~\ref{ta:convert}.

The model Balmer profile has a 
FWHM of 2200 \kms\ and a full width at zero emission of 7700 \kms, whereas the narrow 
bandpasses are generally rectangular in shape with a width of $2500-3000$ \kms.
The conversion factors thus represent a conversion from count rates to total integrated 
emission-line flux, including for \ha\ and \hb\ portions of the broad component that
lie outside the range of the narrow-band filter.  The conversions are therefore 
sensitive to the assumed profile shape.
The model line profile FWHM of 2200 \kms\ is fairly narrow for a 
broad line.  The seventeen radio-quiet QSOs in the $z\sim 2$
\citet{mcea99} sample have a mean \hb\ FWHM of 5100 \kms, 
while the $z < 0.5$ \citet{bogr92} sample has a mean \hb\ FWHM of 3800 \kms.  The
BALQSOs and non-BALQSOs in the \citet{weea91} study have
mean \civ\ half-widths at half maximum of 2200 and 2500 \kms, respectively.
To illustrate the effect the line profile has on our results, we recalculate the
conversion factors for the broad lines assuming a Gaussian profile with a 
FWHM of 5000 \kms.  These conversion factors are listed in column~6 of 
Table~\ref{ta:convert}, and the ratio of these values to those derived using the
\citet{frea91} profile are listed in column~7.  As the table shows, the conversion
factors increase by various amounts up to $\sim 40$\%,
which would result in corresponding increases in the fluxes and equivalent widths
we derive.  The conversions are also sensitive to the position in wavelength space of
the line profile in the narrow-band filter, since for a given line flux the count rate 
will be higher if the line peak is near the center of the bandpass than if it is near 
the edge.  To illustrate this,
we calculate the conversion factor that would be used if the peak of the line were 
exactly at the effective wavelength of the narrow-band filter, using the
\citet{frea91} profile.  These values are listed in column~8 of Table~\ref{ta:convert}.
The ratio of the conversion factors using the true wavelengths to those assuming the 
emission lines are centered are listed in column 9.  It is perhaps useful to consider 
column 9 as a correction factor for the fact that the line is not centered in the 
bandpass.  As one can see by comparing these values to 
the emission-line peak wavelengths and filter effective wavelengths in columns 3 and 4, 
we are correcting significantly more for flux outside the bandpass when the line is 
not centered, a point
which is relevant to the comparison of the BALQSOs and non-BALQSOs in our sample
discussed in \S\ref{sec:discuss}.

The emission-line fluxes, calculated using the conversion factors from column~5 of
Table~\ref{ta:convert}, are
shown in Table~\ref{ta:emline}.  The errors include an assumption of
an uncorrelated 2\%
uncertainty in the absolute throughput of each of the filters, although this makes
a negligible contribution in comparison to the uncertainty in the continuum slope.
(Information on NICMOS photometric calibrations can be found at the NICMOS web site,
\url{http://www.stsci.edu/cgi-bin/nicmos}.)  The errors do not include any
uncertainty in the profile shape.  We also list
in the table the corresponding equivalent widths of the emission lines
in the observer and object rest frames.  It is interesting that of the nine
objects we observed in \ha, the five BALQSOs have systematically larger 
equivalent widths than the four non-BALQSOs.  We discuss this further in
\S\ref{sec:discuss}.

Using aperture photometry to measure the emission-line flux is not 
particularly sensitive as it combines several sources of 
uncertainty, most notably the throughputs of the filters, the QSO continuum 
spectral shape, and the profiles of the emission lines.  As a result, the
only solid detections of any emission-line flux by this method are of \ha\ 
and \hb, most likely dominated by the broad component.  No forbidden lines
typically seen in extended NLRs, such as \oii\ \lamlam 3726, 3729, are 
detected.

Although the observations permit us to measure the strength of the emission
lines by this method, the primary purpose of the observations was to search for
{\it extended} emission.
None of the objects show any obvious extended emission from a simple visual 
inspection of the images (see Figures 1--3).  To probe more sensitively for extended 
narrow-line emission, we need to remove the point-source contribution.  
We experimented with subtracting
scaled broad-band images from the narrow-band images to create a 
continuum-subtracted image as is typically done for extended sources,
but because the emission is (at least mostly) point-like, the resulting
image is dominated by the residuals from the variation in PSFs between
bands.  Instead, we create PSF-subtracted narrow-band
images, using the model PSFs generated individually for each object with
Tiny Tim as discussed previously.  

\begin{figure}[f]
\plotone{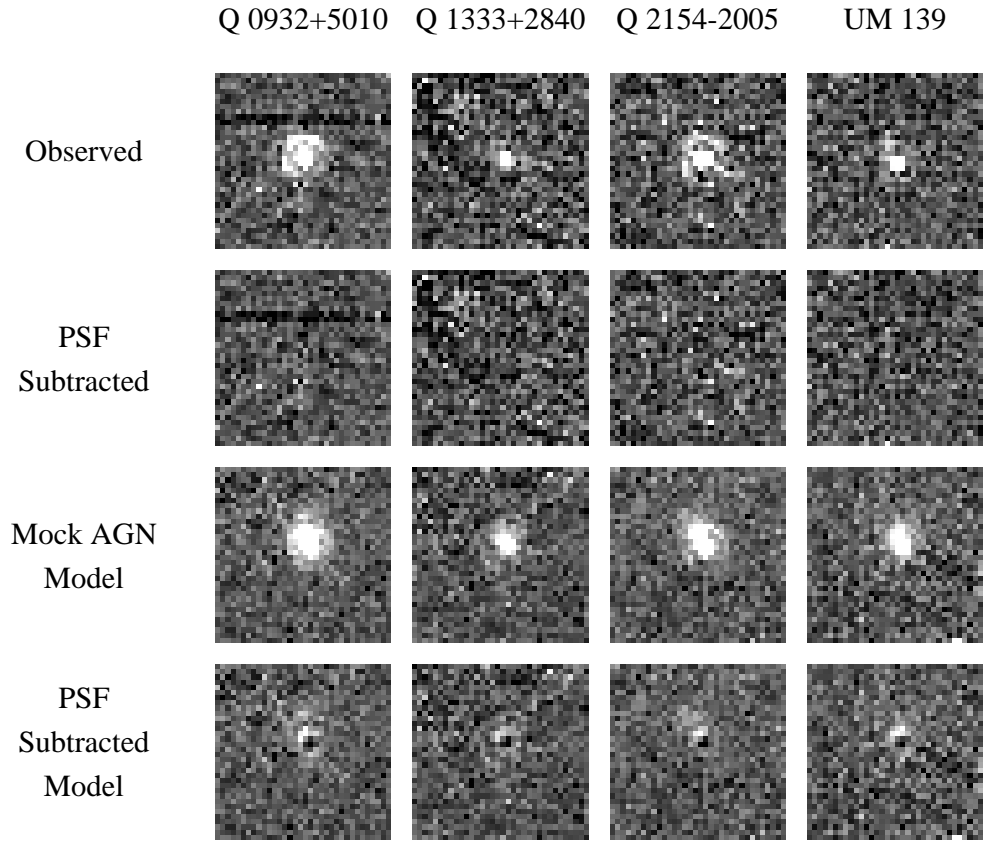}
\caption{The observed narrow band image, the PSF-subtracted narrow-band
image, a mock AGN model with the minimum amount of extended emission that we consider 
to be detectable,
and the PSF-subtracted model for all objects observed with NICMOS camera 1.
Each image is 1.5 $\times$ 1.5 arcsec.\label{fig:nic1}}
\end{figure}

\begin{figure}[f]
\plotone{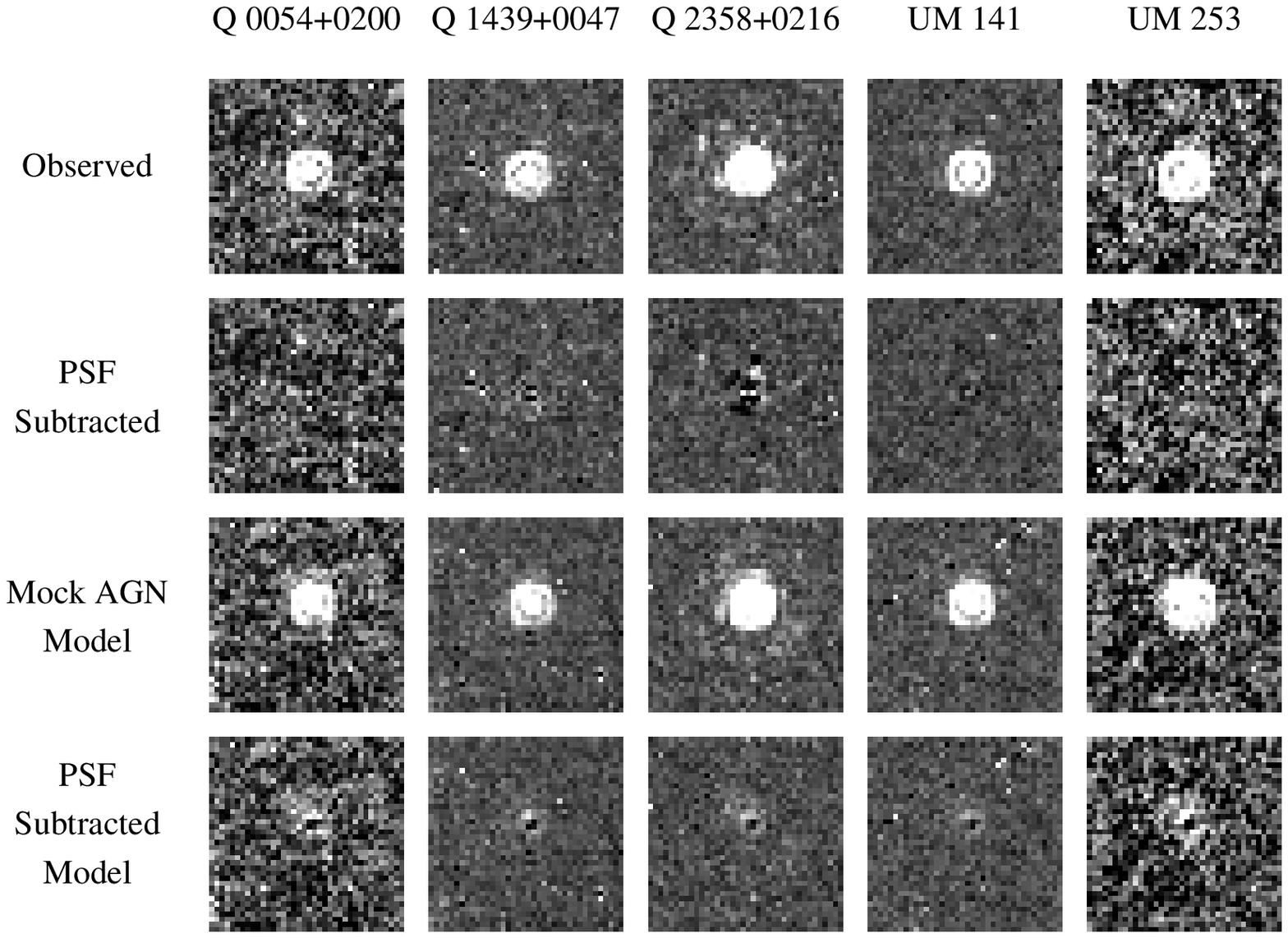}
\caption{Same as Figure~\ref{fig:nic1} for all objects observed with 
NICMOS camera 2.  Each image is 3 $\times$ 3 arcsec.\label{fig:nic2}}
\end{figure}

\begin{figure}[f]
\plotone{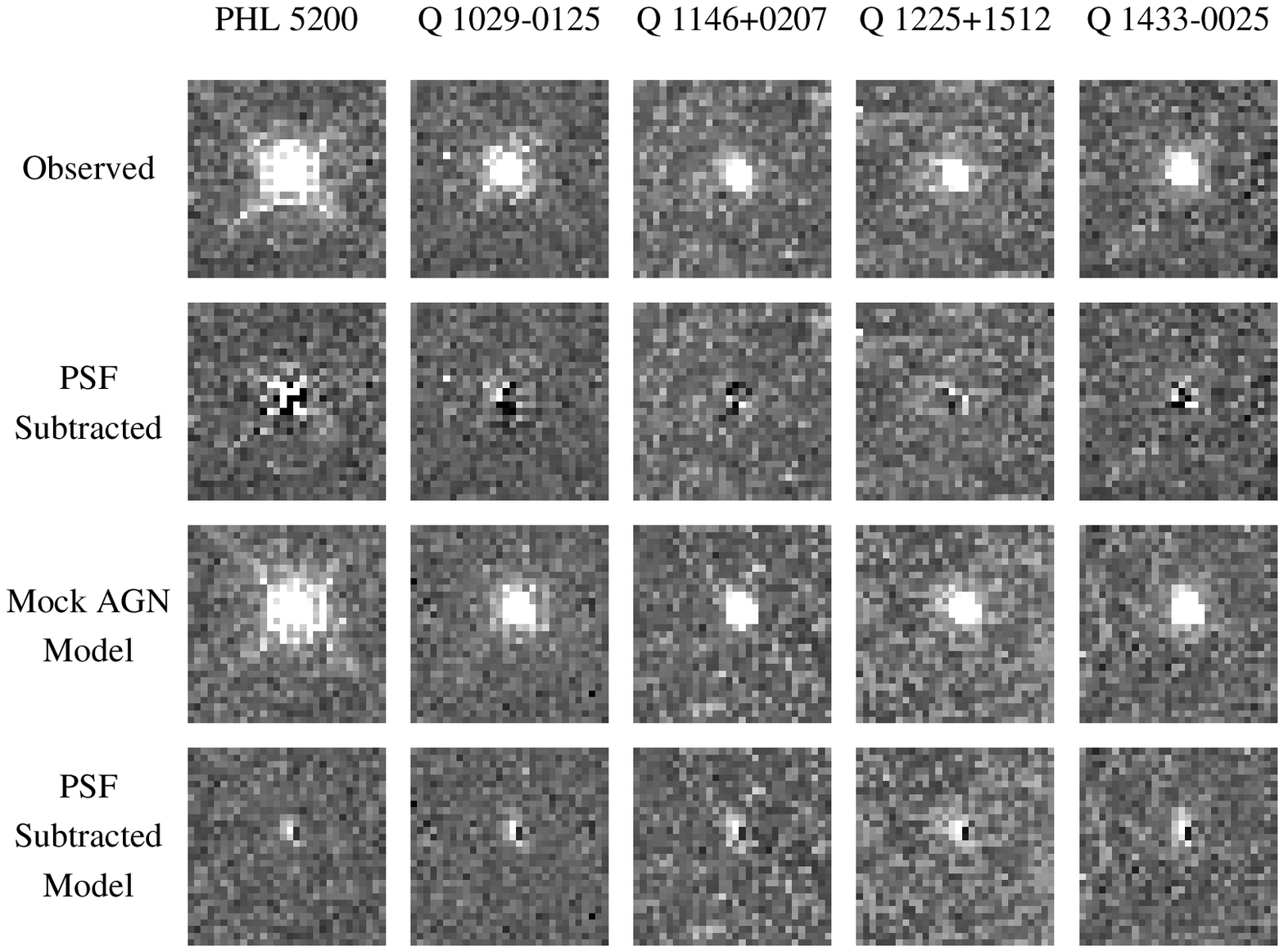}
\caption{Same as Figure~\ref{fig:nic1} and Figure~\ref{fig:nic2} for
all objects observed with NICMOS camera 3.
Each image is 6 $\times$ 6 arcsec.\label{fig:nic3}}
\end{figure}

We use a simple iterative process to find the 
best-fit subtraction by minimizing the RMS deviation of the fit residuals. 
The position and normalization of the point source are varied as free 
parameters in the fitting process.  The PSF subtractions worked quite well in general,
particularly for cameras 1 and 2, although the camera 3 subtractions did leave some 
residuals.  The PSF subtracted images are shown in the second row of 
Figure~\ref{fig:nic1} for the NIC 1 images, Figure~\ref{fig:nic2} for NIC 2, and 
Figure~\ref{fig:nic3} for NIC 3.
We measure the residual count rate after PSF
subtraction in the same fixed aperture as before.  
These count rates are converted to flux, assuming the residual counts are due
to a 400 \kms\ FWHM emission line at the expected wavelength for all lines.  
The results are shown in column 2 of Table~\ref{ta:resid}.
The quoted errors only include the individual pixel noise, 
so they do not include any uncertainty in the
fits or the PSFs.  Since we find residual fluxes both positive and negative well above
this noise level, there are certainly additional sources of uncertainty.
From inspection of the images, we conclude that all of the residual fluxes can
be explained by either imperfect PSF subtractions or by variations in the
background, probably associated with flatfield features not entirely removed
by the pedestal correction.

As a final check on whether any extended emission is present in our images
and to set upper limits on its absence,
we create models to simulate the expected appearance of the extended
emission in the image and the residuals from the PSF subtraction.  
To model the shape of the extended emission we use an appropriately scaled
continuum-subtracted WFPC2 image in \oiii\ of NGC~1068 \citep{dtkf98}.
We assume the same size scaling for the extended emission of each model.
The angular scale at the distance of NGC~1068 is 72 ${\rm pc \ arcsec^{-1}}$
\citep{blea97}.
Assuming $\Omega_0 = 1$ and $H_0 = 100 \ h \ {\rm km \ s^{-1} \ Mpc^{-1}}$, 
the angular scale at $z=2$, approximately the mean redshift of our sample,  
is $4.1 \ h^{-1} \ {\rm kpc \ arcsec^{-1}}$.  For $h=0.75$,
the angular scale is $5.5 \ {\rm kpc \ arcsec^{-1}}$.  To scale NGC~1068 up 
to the luminosity of a QSO, we assume that
the NLR is optically thin and radiation bounded.  Thus
the physical size should scale roughly as the square root of the intensity of 
the ionizing radiation, which should scale roughly with bolometric 
luminosity.  The bolometric luminosity of NGC~1068 is 
$L_{bol, NGC1068} \sim 4.2 \times 10^{44} \ {\rm erg \ s^{-1}}$ \citep{piea94}, 
whereas a typical QSO in our sample has a bolometric luminosity of
$L_{bol, QSO} \sim 2 \times 10^{46} \ {\rm erg \ s^{-1}}$.  We thus expect the
NLR to be $\sim \sqrt{2 \times 10^{46} / 4.2 \times 10^{44}} = 6.9$ times
larger in linear physical extent.  Including the relative angular scales 
due to distance as
described above, the overall size scaling to be applied to the NGC~1068
image to make it resemble a hypothetical BALQSO is $0.09$.  We scale the 
NGC~1068 image by this amount and resample it to match the pixel size of the 
various NICMOS cameras.  We normalize the flux for each model such that the 
total flux of the extended emission is some
particular fraction of the aperture-corrected flux.  We then convolve 
the model with the appropriate PSF and place it onto an empty section 
of the original image.  A point 
source is added to the model, representing the remainder of the flux,
such that the total flux of the point source plus extended component always
equals the aperture-corrected value.  This is then
fit and subtracted using the same method as for the real data.  We repeat
this process for various fractions of extended emission relative to total flux in 
order to find at what minimum fraction the extended emission becomes detectable.
We judge the extended emission to be detectable if some structure
of several contiguous pixels above (or below) the background noise is visually
apparent in the residuals.  This process allows us to place an upper
limit on the observed extended emission, albeit a subjective one.

The models with the minimum detectable fraction of extended emission 
are shown in row 3 of Figures 1--3.
The PSF-subtracted models are shown in row 4 of the figures.  Columns 3 and 4 of 
Table~\ref{ta:resid} list the narrow emission-line fluxes corresponding to the 
minimum detectable flux models and the residuals from the PSF subtractions of these
models.  These residuals were measured in the same fixed aperture as the other
flux measurements and can thus be directly compared to the values from the real data in
column 2.  The residuals from
the models are in most cases clearly larger than what was measured from the real data,
consistent with our conclusion that we do not detect any extended flux.  
It is evident
from comparing columns 3 and 4 of Table~\ref{ta:resid} that the fit requirement of 
minimizing the RMS deviation of the residuals results in a substantial amount of
the extended flux being removed by the PSF subtraction, as much as $\sim 90$\%.

Our tests using the models suggest that the presence of extended emission 
would result in excess flux in the first dark Airy ring as well as an over-subtracted
core in the PSF-subtracted images, as can be seen in row 4 of Figures 1--3.
The camera 1 and 2 data clearly show no such features
in the PSF-subtracted images.  For the only object observed with camera 1 or 2 with any 
significant residuals, Q~2358+0216, the residuals are along the first bright Airy ring 
and in the core.  This is more indicative of an imperfect PSF subtraction than of
extended emission.  The camera 3 data are more difficult to judge.  Although in 
principle camera 3 is most sensitive for our purpose due to the larger pixel
scale and correspondingly lower read noise per unit angular area, the undersampling of 
the PSF makes
the fits less certain and makes it difficult to determine if there is excess flux in 
the first dark Airy ring, which has a width of only $\sim 0.5$ pixels.  Since the
residuals from the camera 3 subtractions do not resemble closely those expected from
the models, we conclude that the residuals are due to problems with the subtraction
and are not indicative of extended emission.

\section{DISCUSSION\label{sec:discuss}}

To provide a context for the limits we have placed on the amount of extended emission,
we use two separate methods to estimate the amount of extended emission we
expected to observe.  For the first estimate we assume that the equivalent widths
of the narrow forbidden emission lines should be the same as for the composite spectrum of
\citet{frea91}.  The equivalent widths for the composite
are 15 \AA\ for \oiii\ \lam 5007 and 1.9 \AA\ for \oii\ \lam 3727.
To estimate narrow-line equivalent widths for the Balmer lines, 
we assume flux ratios of 10 for \oiii\ \lam 5007 / narrow \hb\ and 3 for
narrow \ha\ / narrow \hb.  These ratios are typical for Seyfert 2s and 
narrow-line radio galaxies \citep{ko78}.
Assuming a continuum shape $F_{\nu} \propto \nu ^{-0.3}$, we derive expected
rest-frame equivalent widths of $\sim$7.5 \AA\ for narrow \ha\ and $\sim$1.5 \AA\ 
for narrow \hb.  These equivalent widths are used to estimate narrow-line fluxes for
each object, listed in column 5 of Table~\ref{ta:resid}.

For the second estimate of the expected narrow-line flux we scale the observed
flux of the emission lines in NGC~1068 \citep{ko78}
to the luminosity and distance of the QSOs in our sample.
We assume that the flux of the lines scales linearly with bolometric luminosity.
The bolometric luminosities of the QSOs are estimated as $\lambda F_{\lambda}$ as
observed in
the F160W filter.  The luminosity distance for each QSO is calculated from the
redshift assuming $\Omega = 1$ and $h=0.75$.  The line fluxes scaled from 
NGC~1068 are listed in column 7 of Table~\ref{ta:resid}.

Columns 6 and 8 of Table~\ref{ta:resid} list the ratios of these fluxes to the
minimum detectable flux in column 3 determined from the models.  A ratio greater
than unity would indicate that we would expect to have detected extended emission.
However, only the NGC~1068 scaling for PHL~5200 exceeds unity.  This object had
the poorest PSF subtraction, so it is difficult to ascertain the existence of 
extended emission.  Depending on the scaling, roughly half of the objects are 
above the 10\%
level, while some of the others are much smaller, well below our detection threshold.
If the extended emission is at the level expected by these scaling arguments, it
is not particularly surprising that we did not detect it.  
However, if the emission were only a few times stronger than
what we expect based on these simple scaling arguments, then we should have been
able to detect extended emission in some cases.  

It is unfortunate that none of the six objects on the target list for our snapshot program 
matched for \oiii\ were observed, since the \oiii\ lines are expected to be
the strongest emission lines in the NLR.  For our scaling argument above, we assumed
an \oiii\ \lam 5007 equivalent width of 15 \AA.   \citet{mcea99}
use infrared spectroscopy to study the rest-frame optical emission-line properties of a 
sample of QSOs with similar redshifts and luminosities to our sample.  They find a mean 
equivalent width for \oiii\ \lam 5007 of 8.1 \AA\ for seven BALQSOs and 11.6 \AA\ for 
ten radio-quiet non-BALQSOs, but given the uncertainty and the small sample size the 
difference is not significant.   These equivalent widths are somewhat lower than
the mean of 23 \AA\ for the 87 $z < 0.5$ PG QSOs observed by
\citet{bogr92}.  The low-redshift IR-selected QSO sample of
\citet{bome92} has even higher mean \oiii\ equivalent
width of 46 \AA\ for the 14 objects which were not low-ionization BALQSOs.  Given these
results, with even a conservative estimate of the strength of \oiii, we would expect
to observe \oiii\ line fluxes a factor of 3--5 greater than what we list in 
Table~\ref{ta:resid}.  We therefore might reasonably have expected to see extended emission
in \oiii\ in 1--2 objects had they been observed.

The observed difference in \ha\ equivalent widths between BALQSOs and non-BALQSOs
merits further discussion.  The mean rest-frame \ha\ equivalent width for the
BALQSOs is $390 \pm 30$ \AA, while the mean equivalent width for the non-BALQSOs
is $200 \pm 20$ \AA, corresponding to a ratio 
$\langle{\rm EW}_{H\alpha}{\rm (BAL)}\rangle / \langle{\rm EW}_{H\alpha}{\rm (non-BAL)}\rangle = 1.9 \pm 0.2$.
Even more striking than the ratio of the means is that the difference is systematic ---
{\it all} of the BALQSOs have larger \ha\ equivalent widths than {\it all} of the 
non-BALQSOs.  A two-sample K-S test yields that the BALQSO and non-BALQSO data were
drawn from separate populations with 99.3\%
confidence, although given that the sample groups are small and certainly incomplete,
one should approach this number with caution.  However, the data are
suggestive of a possible real systematic difference between the two groups.

We have considered the possibility that the difference in equivalent widths could
be due to some systematic difference in the observations.  The observations were
performed over roughly the same time period and with the same filters and cameras
for both groups.  However, there is a systematic difference in the location of the
\ha\ line in the filters.  For all of the non-BALQSOs, the expected location of the
peak of the broad \ha\ emission line lies near the center of
the filter bandpass, while for the BALQSOs the peak of the line lies near the edge of 
the bandpass, resulting in larger corrections for flux outside the bandpass.
This can be seen by comparing the values in column~9 of Table~\ref{ta:convert} for
the BALQSOs and non-BALQSOs.
For three of the BALQSOs the emission line center is longward 
of the center of the filter, while for the other two it is shortward.
The systematic difference in the location of the lines in the filters renders the effect
of making the relative \ha\ equivalent widths of the BALQSOs and non-BALQSOs sensitive to
the assumed profile shape.  We find that if we assume a broader line profile, the 
disparity between the two groups diminishes but does not disappear.
If we use the conversion factors from column 6 of Table~\ref{ta:convert}, calculated
assuming a Gaussian profile with a FWHM of 5000 \kms, the ratio of equivalent widths
becomes
$\langle{\rm EW}_{H\alpha}{\rm (BAL)}\rangle / \langle{\rm EW}_{H\alpha}{\rm (non-BAL)}\rangle = 1.6 \pm 0.2$.

There are several possible systematic differences between BALQSOs and non-BALQSOs
that could explain our observations.  The most straightforward explanation is that
there is simply more \ha\ flux in BALQSOs than non-BALQSOs.  Another possibility is
that the \ha\ lines could have the same flux on average but 
be systematically much broader in the non-BALQSOs.  Either one of these would be
extremely surprising, however, since broad emission lines in the UV exhibit no
large systematic differences \citep{weea91}.  Also,
\citet{mcea99} do not find large differences in \hb\ emission 
between BALQSOs and non-BALQSOs.  An even more 
intriguing possibility is that the continua are different in BALQSOs and non-BALQSOs,
since in all cases we have estimated the continuum in the narrow-band filter from
shorter wavelength F160W images.  If the continuum shape were redder than we
have assumed in BALQSOs, we would underestimate the continuum in
the narrow-band images of the BALQSOs, resulting in larger fluxes and equivalent
widths.  The implied continuum shape in the rest-frame optical for BALQSOs that would
account for the systematically larger \ha\ equivalent widths has an
effective power-law index $\alpha_{\nu} = -1.9$, compared to the typical
$\alpha_{\nu} = -0.3$ that we assumed.  This amount of reddening is comparable
with that observed in low-ionization BALQSOs \citep{weea91} and 
attributed to intrinsic dust extinction \citep{spfo92}.
In fact, two of the BALQSOs in our sample, PHL~5200 and Q~2358+0216, are
low-ionization BALQSOs, so it is not surprising that we derive larger \ha\ 
equivalent widths for these objects.  However, the fact that the other three
BALQSOs follow the same trend is interesting, since there is generally little
evidence from rest-frame UV spectra for intrinsic dust extinction in 
high-ionization BALQSOs \citep{weea91}.
To investigate whether there is an overall reddening of the continuum in our sample,
we used our fluxes in the F160W band for each object along with the fluxes at 
$\sim 2100$ \AA\ in the rest frame calculated from optical spectra by 
\citet{weea91} to derive $(R-H)$ color indices.  For the two
QSOs in our sample not in \citet{weea91}, Q~0054+0200 and
Q~2358+0216, we estimate the optical flux from the $B_J$ magnitudes in the LBQS catalog 
\citep{hfc95}.  Of the objects observed in \ha, we find that
both the BALQSO and non-BALQSO groups have $\langle (R-H) \rangle = 1.95$ with
$\sigma \sim 0.8 \ {\rm mag}$, suggesting that there is no significant systematic 
difference in the 
continuum properties in our sample in the rest-frame 2100--5500 \AA\ spectral region. 
Near-infrared (rest-frame optical) spectroscopy of BALQSOs is 
necessary to resolve this issue.

\section{SUMMARY\label{sec:summary}}

We have obtained {\it HST} NICMOS snapshot images of ten BALQSOs and four non-BALQSOs
to search for extended narrow-line emission.  Each object was imaged in both the
F160W filter and a narrow-band filter corresponding to \ha, \hb, or \oii.  Using
aperture photometry, we derive estimates of the emission-line fluxes and 
equivalent widths.  To look for extended emission, we use model PSFs to generate
PSF-subtracted narrow-band images.  We measure the residual flux in these images
and compared them to PSF-subtracted images of simulated data.  Our most important 
results are:
\begin{enumerate}
\item{We do not detect extended emission in any of the images.  Some of the images
have PSF-subtraction residuals, but based on comparisons with simulated data
we conclude that the residuals are the result of imperfect PSF subtraction and
are not evidence for extended emission.}
\item{We find that for the nine objects we observed in \ha, we derive systematically
larger equivalent widths for the five BALQSOs than for the four non-BALQSOs, by a 
factor of 1.9 on average.  Given the small sample size, 
it is not certain whether this is a real effect or just a statistical aberration.
If it is real, it could be evidence either for an intrinsic difference between BALQSOs
and non-BALQSOs in \ha\ emission, or for systematically redder rest-frame optical
continua in BALQSOs, an effect which has been observed in low-ionization BALQSOs.
If either of these differences were real, it would stand in stark
contrast to studies of rest-frame UV spectra which find little 
difference between the continuum and emission-line properties of high-ionization
BALQSOs and radio-quiet non-BALQSOs.}
\end{enumerate}

\acknowledgments
We are grateful to D. Calzetti and H. Bushouse for their help in understanding
the NICMOS error arrays.
Support for this work was provided by NASA through
grant number GO-07892.02-96A from the Space Telescope Science Institute,
which is operated by the Association of Universities for Research in Astronomy,
Inc., under NASA contract NAS5-26555.
G. Kriss and Z. Tsvetanov acknowledge additional support from NASA Long Term
Space Astrophysics grant NAGW-4443 to the Johns Hopkins University.

\newpage
\begin{deluxetable}{lcclccc}
\tablecaption{Observations\label{ta:observations}}
\tablehead{
			&					&
			&					&
			&		\colhead{Broad-Band}	&
\colhead{Narrow-Band}\\
\colhead{Object}	&		\colhead{Redshift}	&
\colhead{BAL?}		&		\colhead{Date}		&
\colhead{Camera}	&		\colhead{Filter}	&
\colhead{Filter}}
\startdata
PHL 5200	& 1.981 & Y & 1998 June 16       & 3 & F160W & F196N\\
Q 0021-0213\tablenotemark{*}	& 2.348 & Y & 1998 October 10    & 1 & F160W & F164N\\
Q 0054+0200	& 1.872 & Y & 1998 August 3      & 2 & F160W & F187N\\
Q 0932+5010	& 1.914 & Y & 1998 September 13  & 1 & F160W & F108N\\
Q 1029-0125	& 2.029 & Y & 1998 June 20       & 3 & F160W & F200N\\
Q 1146+0207	& 2.054 & N & 1998 March 15      & 3 & F160W & F200N\\
Q 1225+1512	& 1.990 & N & 1998 June 14       & 3 & F160W & F196N\\
Q 1333+2840     & 1.908 & Y & 1998 November 10   & 1 & F160W & F108N\\
Q 1433-0025	& 2.042 & N & 1998 May 8         & 3 & F160W & F200N\\
Q 1439+0047     & 1.857 & N & 1998 July 10       & 2 & F160W & F187N\\
Q 2154-2005	& 2.035 & Y & 1998 October 26    & 1 & F160W & F113N\\
Q 2358+0216	& 1.872 & Y & 1998 August 16     & 2 & F160W & F187N\\
UM 139		& 2.029 & Y & 1998 October 18    & 1 & F160W & F113N\\
UM 141		& 2.909 & Y & 1998 September 27  & 2 & F160W & F190N\\
UM 253		& 2.253 & Y & 1998 October 31    & 2 & F160W & F212N\\
\enddata
\tablenotetext{*}{Because of an incorrect redshift, no emission line is in the F164N
bandpass, so we ignore this object in our analysis.}
\end{deluxetable}

\begin{deluxetable}{lccccc}
\tablecaption{Aperture Photometry\label{ta:fluxes}}
\tablehead{
\colhead{}		&\multicolumn{2}{c}{Broad-Band Count Rates\tablenotemark{*}} &
\multicolumn{2}{c}{Narrow-Band Count Rates\tablenotemark{*}}	& \colhead{} \\
\cline{2-3} \cline{4-5} \\
\colhead{Object}	&		\colhead{Measured}	&
\colhead{Corrected}	&		\colhead{Measured}	&
\colhead{Corrected}	&		\colhead{$AB_{F160W}$}}
\startdata
PHL 5200	& $249.7 \pm 0.8$	& $258.4 \pm 0.8$	&
		  $62.9 \pm 0.1$	& $71.8 \pm 0.1$	&
		  $16.821 \pm 0.003$	\\
Q 0021-0213	& $105.1 \pm 3.1$	& $115.8 \pm 3.4$	&
		  $5.1 \pm 0.3$		& $5.6 \pm 0.4$		&
		  $17.793 \pm 0.032$	\\
Q 0054+0200	& $75.9 \pm 1.4$	& $84.4 \pm 1.5$	&
		  $13.2 \pm 0.2$	& $14.6 \pm 0.2$	&
		  $18.294 \pm 0.019$	\\
Q 0932+5010	& $273.8 \pm 2.5$	& $301.6 \pm 2.7$	&
		  $4.1 \pm 0.2$		& $4.5 \pm 0.2$		&
		  $16.754 \pm 0.010$	\\
Q 1029-0125	& $113.0 \pm 0.7$	& $116.9 \pm 0.7$	&
		  $22.1 \pm 0.1$	& $25.3 \pm 0.1$	&
		  $17.682 \pm 0.006$	\\
Q 1146+0207	& $123.3 \pm 0.7$	& $127.6 \pm 0.7$	&
		  $14.8 \pm 0.1$	& $17.0 \pm 0.1$	&
		  $17.505 \pm 0.006$	\\
Q 1225+1512	& $74.4 \pm 1.2$	& $77.0 \pm 1.2$	&
		  $16.4 \pm 0.1$	& $18.7 \pm 0.1$	&
		  $18.136 \pm 0.017$	\\
Q 1333+2840	& $96.7 \pm 3.5$	& $106.5 \pm 3.8$	&
		  $1.5 \pm 0.3$		& $1.6 \pm 0.3$		&
		  $17.884 \pm 0.039$	\\
Q 1433-0025	& $64.8 \pm 0.6$	& $67.1 \pm 0.6$	&
		  $11.0 \pm 0.1$	& $12.1 \pm 0.1$	&
		  $18.204 \pm 0.010$	\\
Q 1439+0047	& $68.1 \pm 1.5$	& $75.7 \pm 1.7$	&
		  $10.7 \pm 0.2$	& $12.1 \pm 0.2$	&
		  $18.413 \pm 0.024$	\\
Q 2154-2005	& $155.2 \pm 3.2$	& $171.0 \pm 3.5$	&
		  $3.9 \pm 0.3$		& $4.2 \pm 0.3$		&
		  $17.370 \pm 0.022$	\\
Q 2358+0216	& $214.7 \pm 1.4$	& $238.7 \pm 1.5$	&
		  $35.2 \pm 0.2$	& $39.0 \pm 0.2$	&
		  $17.165 \pm 0.007$	\\
UM 139		& $78.5 \pm 3.0$	& $86.5 \pm 3.3$	&
		  $1.3 \pm 0.3$		& $1.4 \pm 0.3$		&
		  $18.110 \pm 0.041$	\\
UM 141		& $207.4 \pm 1.4$	& $230.6 \pm 1.5$	&
		  $13.8 \pm 0.1$	& $15.3 \pm 0.2$	&
		  $17.203 \pm 0.007$	\\
UM 253		& $97.4 \pm 1.4$	& $108.3 \pm 1.5$	&
		  $21.6 \pm 0.2$	& $23.8 \pm 0.2$	&
		  $18.023 \pm 0.015$	\\
\enddata
\tablenotetext{*}{DN s$^{-1}$}
\end{deluxetable}

\begin{deluxetable}{lcccccccc}
\tablecaption{Emission Line Flux Conversion Factors\label{ta:convert}}
\tablehead{
\colhead{}		&	\colhead{Emission}	&
\colhead{Redshifted}	&	\colhead{Filter} &
\colhead{}		&	\colhead{}	&
\colhead{}		&	\colhead{}	&
\colhead{} \\
\colhead{Object}	&	\colhead{Line}	&
\colhead{$\lambda _{peak}$(\AA)}	&
\colhead{$\lambda _{eff}$(\AA)} 	&
\colhead{A}		&	\colhead{B} &
\colhead{B/A}		&	\colhead{C} &
\colhead{A/C} \\
\colhead{(1)}		&	\colhead{(2)}		&
\colhead{(3)}		&	\colhead{(4)}		&
\colhead{(5)}		&	\colhead{(6)}		&
\colhead{(7)}		&	\colhead{(8)}		&
\colhead{(9)}}
\startdata
PHL 5200	& \ha	& 19569 & 19639 &
		1.986 	& 2.430	& 1.22	& 1.017 & 1.95 \\
Q 0054+0200	& \ha	& 18854 & 18740 & 
		2.976	& 3.259 & 1.10	& 1.805 & 1.65 \\
Q 0932+5010	& \oii	& 10865 & 10816 &
		13.40	& \nodata & \nodata & 7.592 & 1.77 \\
Q 1029-0125	& \ha	& 19884 & 19975 &
		2.147	& 2.480 & 1.10	& 1.584 & 1.36 \\
Q 1146+0207	& \ha	& 20048 & 19975 &
		1.823	& 2.354 & 1.29	& 1.584 & 1.15 \\
Q 1225+1512	& \ha	& 19628 & 19639 &
		1.621	& 2.193 & 1.35	& 1.017 & 1.59 \\
Q 1333+2840	& \oii	& 10842 & 10816 & 
		7.665	& \nodata & \nodata & 7.592 & 1.01 \\
Q 1433-0025	& \ha	& 19970 & 19975 &
		1.589	& 2.111 & 1.33	& 1.584	& 1.00 \\
Q 1439+0047	& \ha	& 18755 & 18740 &
		1.807	& 2.438 & 1.35	& 1.805 & 1.00 \\
Q 2154-2005	& \oii	& 11316 & 11298 &
		5.880	& \nodata & \nodata & 6.110 & 0.96 \\
Q 2358+0216	& \ha	& 18854 & 18740 &
		2.976	& 3.259	& 1.10	& 1.805	& 1.65 \\
UM 139		& \oii	& 11294 & 11298 &
		6.279	& \nodata & \nodata & 6.110 & 1.03 \\
UM 141		& \hb	& 19008 & 19004 &
		1.760	& 2.420 & 1.38 	& 1.759	& 1.00 \\
UM 253		& \ha	& 21355	& 21213 &
		2.722	& 2.757	& 1.01	& 1.391	& 1.96 \\
\enddata
\tablenotetext{~}{Col.\ (3)---Expected wavelength in vacuum of the peak of the redshifted
emission line.}
\tablenotetext{~}{Col.\ (4)---Effective wavelength of the narrow band filter.}
\tablenotetext{~}{Col.\ (5)---Case A:  Conversion factors from count rate to integrated 
emission-line flux assuming a 400 \kms\ FWHM Gaussian for the narrow lines and the \hb\ 
profile extracted from \citet{frea91} for the broad lines.  Units are 
$10^{-16}$ \fluxconv.}
\tablenotetext{~}{Col.\ (6)---Case B:  For the broad lines only, conversion factors
assuming a 5000 \kms\ FWHM Gaussian.  Units are $10^{-16}$ \fluxconv.}
\tablenotetext{~}{Col.\ (7)---Ratio of B to A.}
\tablenotetext{~}{Col.\ (8)---Case C:  Conversion factors that would be used if the 
emission lines were centered in the narrow band filters, assuming the same profiles as A.
Units are $10^{-16}$ \fluxconv.}
\tablenotetext{~}{Col.\ (9)---Ratio of A to C.} 
\end{deluxetable}

\begin{deluxetable}{lcccccc}
\tablecaption{Emission Line Data\label{ta:emline}}
\tablehead{
\colhead{}	& \colhead{}					&
\multicolumn{2}{c}{Narrow-Band Count Rates\tablenotemark{a}} &
\colhead{}	& \multicolumn{2}{c}{Equivalent Width (\AA)} \\
\cline{3-4} \cline{6-7}
\colhead{}		& \colhead{}					&
\colhead{Scaled}	&
\colhead{Net}		& \colhead{Integrated}				&
\colhead{}		& \colhead{} \\
\colhead{Object}		&	\colhead{Line}			&
\colhead{Continuum}		&
\colhead{Emission Line}	&
\colhead{Line Flux\tablenotemark{b}}			&
\colhead{Observed}		&	\colhead{Rest Frame}}
\startdata
PHL 5200	&\ha		&
$15.2^{+1.8}_{-2.0}$		& $56.6^{+2.0}_{-1.8}$		&
$113^{+5}_{-5}$ 		&
$1300^{+220}_{-200}$		& $438^{+74}_{-65}$\\
Q 0054+0200	&\ha		&
$4.42^{+0.42}_{-0.47}$		& $10.19^{+0.50}_{-0.45}$		&
$30.3^{+1.7}_{-1.6}$ 		&
$1050^{+160}_{-140}$		& $364^{+56}_{-50}$\\
Q 0932+5010	&\oii		&
$4.3^{+0.9}_{-1.2}$		& $0.2^{+1.2}_{-0.9}$			&
$3^{+16}_{-13}$ 		&
$8^{+54}_{-43}$			& $3^{+18}_{-15}$\\
Q 1029-0125	&\ha		&
$7.2^{+0.9}_{-1.0}$		& $18.1^{+1.1}_{-0.9}$			&
$38.9^{+2.5}_{-2.2}$ 		&
$1030^{+210}_{-180}$		& $339^{+69}_{-60}$\\
Q 1146+0207	&\ha		&
$7.8^{+1.0}_{-1.1}$		& $9.2^{+1.1}_{-1.0}$			&
$16.7^{+2.2}_{-1.9}$ 		&
$400^{+110}_{-90}$		& $133^{+36}_{-31}$\\
Q 1225+1512	&\ha		&
$4.54^{+0.54}_{-0.61}$		& $14.14^{+0.62}_{-0.54}$		&
$22.9^{+1.2}_{-1.1}$ 		&
$890^{+160}_{-140}$		& $299^{+53}_{-47}$\\
Q 1333+2840	&\oii		&
$1.53^{+0.33}_{-0.41}$		& $0.10^{+0.50}_{-0.44}$		&
$0.8^{+3.9}_{-3.3}		$ &
$7^{+38}_{-32}$			& $3^{+13}_{-11}$\\
Q 1433-0025	&\ha		&
$4.10^{+0.52}_{-0.60}$		& $7.96^{+0.60}_{-0.52}$		&
$12.7^{+1.0}_{-0.9}$ 		&
$580^{+130}_{-110}$		& $191^{+42}_{-37}$\\
Q 1439+0047	&\ha		&
$3.96^{+0.38}_{-0.42}$		& $7.87^{+0.45}_{-0.42}$		&
$14.2^{+0.9}_{-0.9}$ 		&
$550^{+90}_{-80}$		& $191^{+31}_{-28}$\\
Q 2154-2005	&\oii		&
$3.03^{+0.58}_{-0.72}$		& $1.19^{+0.77}_{-0.65}$		&
$7.0^{+4.6}_{-3.8}$ 		&
$44^{+38}_{-32}$		& $14^{+13}_{-10}$\\
Q 2358+0216	&\ha		&
$12.5^{+1.2}_{-1.3}$		& $26.5^{+1.3}_{-1.2}$			&
$79^{+5}_{-4}$	 		&
$960^{+150}_{-130}$		& $335^{+52}_{-46}$\\
UM 139		&\oii		&
$1.53^{+0.30}_{-0.37}$		& $-0.17^{+0.46}_{-0.40}$		&
$-1.1^{+2.9}_{-2.5}$ 	&
$-14^{+33}_{-29}$		& $-4^{+11}_{-10}$\\
UM 141		&\hb		&
$11.9^{+1.2}_{-1.3}$		& $3.5^{+1.4}_{-1.2}$			&
$6.1^{+2.4}_{-2.2}$ 		&
$79^{+40}_{-36}$		& $20^{+11}_{-9}$\\
UM 253		&\ha		&
$6.7^{+1.1}_{-1.3}$		& $17.2^{+1.3}_{-1.2}$			&
$46.8^{+3.7}_{-3.2}$ 		&
$1550^{+410}_{-340}$		& $480^{+120}_{-100}$\\
\enddata
\tablenotetext{a}{DN s$^{-1}$}
\tablenotetext{b}{$10^{-16}\ {\rm erg \ cm^{-2} \ s^{-1}}$}
\end{deluxetable}

\begin{deluxetable}{lccccccc}
\tabletypesize{\footnotesize}
\tablecaption{PSF Subtraction Residuals\label{ta:resid}}
\tablehead{
\colhead{}			&	\colhead{}		&
\colhead{}			&	\colhead{}		&
\multicolumn{2}{c}{Composite QSO Scaling}	&
\multicolumn{2}{c}{NGC 1068 Scaling} \\
\cline{5-6} \cline{7-8}
\colhead{}			&	\colhead{Residual}	&
\colhead{Minimum}		&	\colhead{Model} 	&
\colhead{Expected}		&	\colhead{}		&
\colhead{Expected}		&	\colhead{} \\
\colhead{Object}		&	\colhead{Flux\tablenotemark{a}}		&
\colhead{Detectable Flux\tablenotemark{a}}	&	
\colhead{Residual Flux\tablenotemark{a}}	&
\colhead{Line Flux\tablenotemark{a}}		&
\colhead{Ratio\tablenotemark{b}}		&
\colhead{Line Flux\tablenotemark{a}}		&
\colhead{Ratio\tablenotemark{b}}}
\startdata
PHL 5200	& $2.8 \pm 0.1$				&
4.9 & 0.4 & 1.9 & 0.39 & 6.0 & 1.21\\
Q 0054+0200	& $-0.5 \pm 1.1$				&
33 &  7.7 & 0.6 & 0.02 & 1.5 & 0.05\\
Q 0932+5010	& $-4.8 \pm 2.7$			&
30 & 19.6 & 1.7 & 0.06 & 1.1 & 0.04\\
Q 1029-0125	& $0.1 \pm 0.1$				&
6 & 0.6 & 0.9 & 0.16 & 2.7 & 0.49\\
Q 1146+0207	& $-0.1 \pm 0.1$			&
10 & 1.1 & 0.9 & 0.09 & 3.2 & 0.30\\
Q 1225+1512	& $0.5 \pm 0.1$				&
13 & 1.3 & 0.6 & 0.05 & 1.8 & 0.14\\
Q 1333+2840	& $1.8 \pm 2.3$				&
16 & 5.1 & 0.6 & 0.04 & 0.4 & 0.02\\
Q 1433-0025	& $0.5 \pm 0.1$				&
7 & 0.6 & 0.5 & 0.07 & 1.7 & 0.23\\
Q 1439+0047	& $0.9 \pm 0.3$				&
5 & 0.9 & 0.6 & 0.12 & 1.4 & 0.30\\
Q 2154-2005	& $4.5 \pm 1.8$				&
15 & 6.2 & 0.9 & 0.06 & 0.6 & 0.04\\
Q 2358+0216	& $-2.2 \pm 1.1$			&
22 & 6.4 & 1.8 & 0.08 & 4.4 & 0.20\\
UM 139		& $-1.6 \pm 1.9$			&
16 & 6.2 & 0.5 & 0.03 & 0.3 & 0.02\\
UM 141		& $-0.4 \pm 0.1$			&
4 & 0.7 & 0.5 & 0.12 & 0.9 & 0.25\\
UM 253		& $-7.3 \pm 1.9$			&
114 & 29.2 & 0.7 & 0.01 & 2.0 & 0.02\\
\enddata
\tablenotetext{a}{$10^{-16}\ {\rm erg \ cm^{-2} \ s^{-1}}$}
\tablenotetext{b}{Expected flux / minimum detectable flux}
\end{deluxetable}

\newpage
\epsscale{0.7}

\end{document}